\documentclass[aps,prb,reprint,superscriptaddress,nobibnotes,showkeys,tightenlines]{revtex4-1}
\usepackage{amsmath}
\usepackage{bbm}
\usepackage{commath}
\usepackage{graphicx}
\usepackage{color}
\usepackage{indentfirst}
\usepackage{setspace}
\usepackage{tabularx}
\usepackage{multirow}
\usepackage{subfigure}
\usepackage{float}
\usepackage{ragged2e}
\usepackage{natbib}
\usepackage{titlesec}
\titlespacing*{\section}{0pt}{0.5\baselineskip}{0.5\baselineskip}

\begin{document}

\title{Melting temperature prediction via first principles and deep learning} 

\author{Qi-Jun Hong}
\email[e-mail:]{ qhong@alumni.caltech.edu}
\affiliation{School for Engineering of Transport, Energy and Matter, Arizona State University, Tempe, AZ 85287, USA}

\date{\today}

\begin{abstract}

Melting is a high temperature process that requires extensive sampling of configuration space, thus making melting temperature prediction computationally very expensive and challenging.
Over the past few years, I have built two methods to address this challenge, one via direct density functional theory (DFT) molecular dynamics (MD) simulations and the other via deep learning graph neural networks.
The DFT approach is based on statistical analysis of small-size solid-liquid coexistence MD simulations. It eliminates the risk of metastable superheated solid in the fast-heating method, while also significantly reducing the computer cost relative to the traditional large-scale coexistence method. Being both accurate and efficient (at the speed of several days per material), it is considered as one of the best methods for direct DFT melting temperature calculation.
The deep learning method is based on graph neural networks that effectively handles permutation invariance in chemical formula, which drastically improves  efficiency and reduces cost. At the speed of milliseconds per material, the model is extremely fast, while being moderately accurate, especially within the composition space expanded by the dataset.
I have implemented both methods into automated computer code packages, making them publicly available and free to download.
The DFT and deep learning methods are highly complementary to each other, and hence they can be potentially well integrated into a framework for melting temperature prediction.
I demonstrated examples of applying the methods to materials design and discovery of high-melting-point materials.
\end{abstract}
\maketitle

Melting temperature is an important materials property, especially in phase diagram construction and in the design and discovery of high-performance refractory materials, which serve crucial roles in applications ranging from gas turbine engines to heat shields for hypersonic vehicles \cite{Padture2002,Wuchina2007,Perepezko2009,Lu2010,Liu2013,Padture2016}.
Furthermore, high melting temperature often correlates with desirable thermodynamic and mechanical properties, such as high-temperature materials strength, good ablation, and creep resistance.
However, unlike certain materials properties that can be readily obtained through simple static calculations at absolute zero, melting temperature is a high temperature property that involves an expensive and complex computation procedure, especially since it requires modeling of the liquid phase and thus a large amount of configuration sampling. As a result, melting temperature calculation is considered very challenging and our capabilities are still limited throughout the computational community.

Numerous ingenious methods have been devised in attempt to capture melting temperatures from computation \cite{Hongthesis}.
Using empirical potentials is relatively inexpensive, but it depends on availability and reliability of the empirical potentials. It is both complicated and time-consuming to building a new classical interatomic potential for every new material, not to mention the issue of reliability regarding accuracy.
Density functional theory (DFT) calculations are clearly better in terms of generalizability and reliability. However, they remain notoriously expensive, despite increasing power and capability of our computers.
The large-size coexistence method \cite{Mei1992, Morris1994}, which is generally considered the gold-standard and widely utilized as a benchmark, typically requires a system size too large for DFT simulations, rendering this approach prohibitively expensive in practice.
The single-phase small-size Z method \cite{Belonoshko2006}, which heats a solid until it melts, suffers from hysteresis. Despite proposed solutions based on homogeneous melting to alleviate superheating, this approach still lacks a rigorous physical foundation \cite{Alfe11}.
Alternatively, one can compute melting temperatures via the free energy method \cite{Mei1992, Sugino1995}, which locates the intersection of the free energy curves of the solid and the liquid. This approach requires highly accurate free energy calculation of the liquid phase, because the two curves cross at a terribly shallow angle and thus a small free energy shift will result in a large error in melting temperature. Unfortunately, all methods for liquid-state free energy computation are expensive and challenging, such as thermodynamic integration \cite{deWijs98}, the particle insertion method \cite{Widom1982, Hong2012}, and the two-phase thermodynamics method \cite{Lin2003}.

In this work, I review the two methods we have built over the past few years, to achieve the goal of melting temperature prediction.
In the first method, I built an accurate and robust first principles approach, together with Dr. Axel van de Walle.
The method, called small-size coexistence, \cite{Hong2013,Hongthesis}, is highly accurate and efficient, with an error generally around 100K and a cost of several days per material. The DFT approach is based on statistical analysis of small-size solid-liquid coexistence MD simulations. It eliminates the risk of metastable superheated solid in the fast-heating method, while also significantly reducing the computer cost relative to the traditional large-scale coexistence method. I have also further automated the computational process into the \textit{SLUSCHI} (Solid and Liquid in Ultra Small Coexistence with Hovering Interfaces) package \cite{Hong2016}, which I have freely distributed and is publicly available at my group's website. I have utilized this method to study hundreds of materials \cite{HfCN2015,Hong2015, Ta_EOS2015, Kapush2017, Hong2018,Hong2019, Ushakov2019, Fyhrie2019, Hong2021}.
In the second method, I have built an extremely fast machine learning model, via deep learning, to predict melting temperature from chemical formula. 
The method is based on graph neural networks that effectively handles permutation invariance in chemical formula (e.g., NaCl is equivalent to ClNa), which drastically improves efficiency and reduces cost.
At the speed of milliseconds per material, the model is extremely rapid, while being moderately accurate, particularly within the composition space expanded by the dataset. I have deployed the model into production, which is publicly available at our group's webpage. It is very easy to run the model, and a user needs to type in only the chemical formula in order to generate an estimate of its melting temperature in seconds.

%\begin{figure*}[t]
%\includegraphics[width=1.05\textwidth]{Picture1.png}
%\caption{\label{database}A screenshot of the melting temperature database we have built so far. The database currently contains 9375 materials, out of which 982 compounds are high-melting-temperature materials with melting points above 2000K.}
%\end{figure*}
\begin{figure*}
\includegraphics[width=0.45\textwidth]{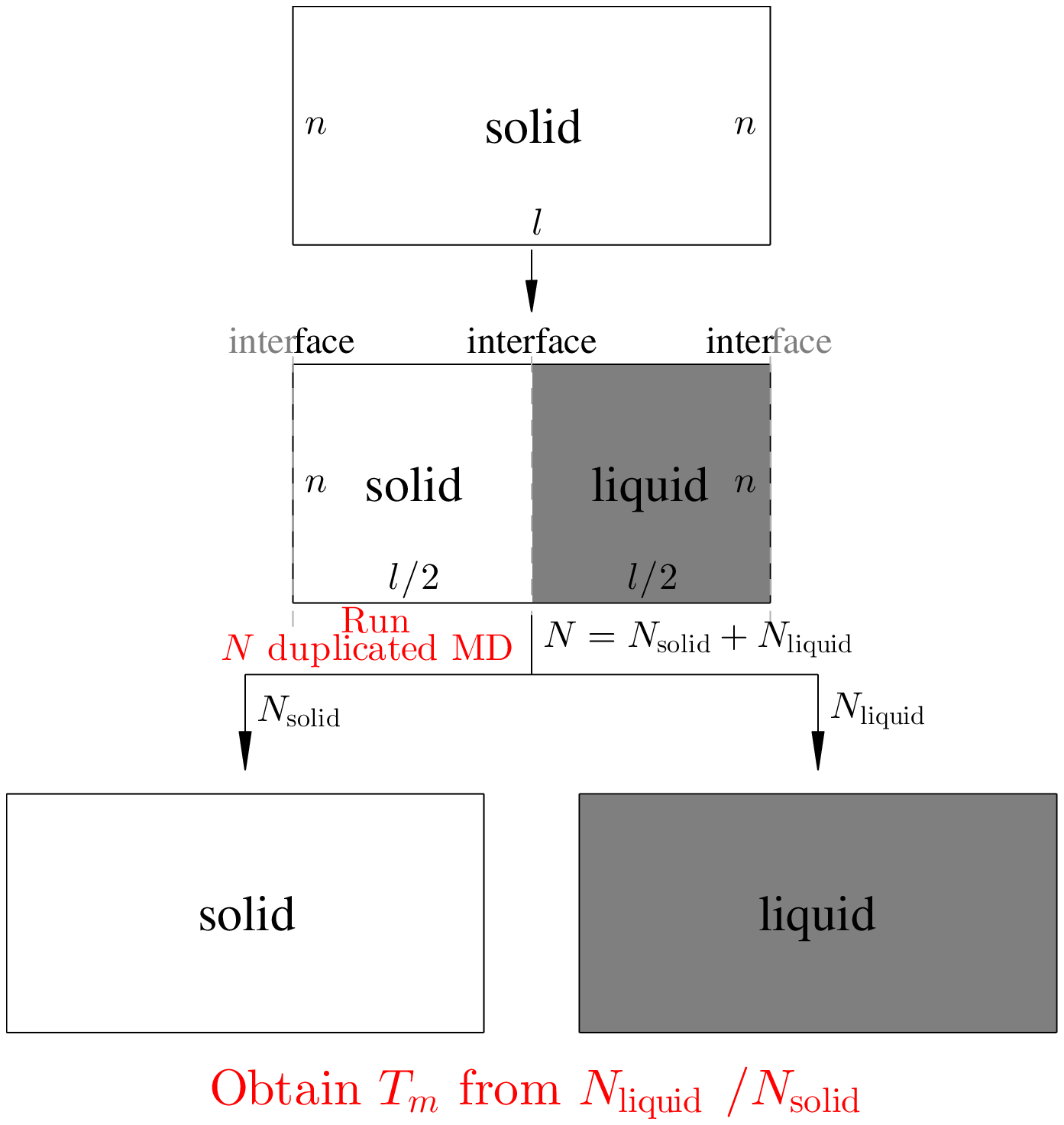} \hspace{.2cm}
\includegraphics[width=0.49\textwidth]{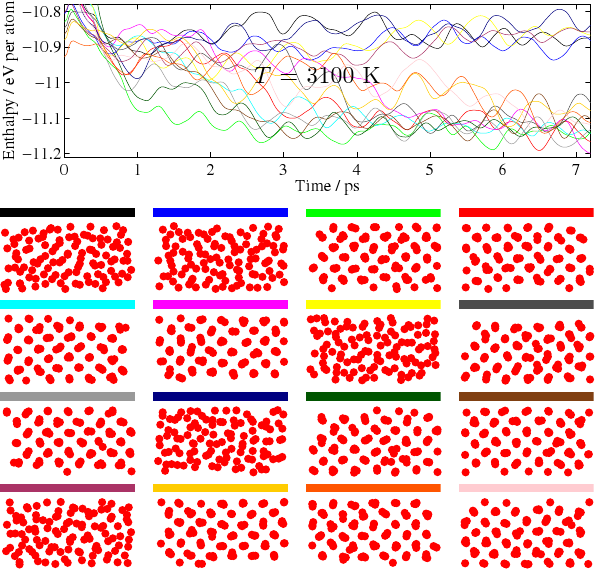}
\caption{\label{diagram}Left: Schematic illustration of how small-size coexistence method is executed in practice. Starting from $n \times n \times l$ supercell with atoms at their ideal solid positions, I heat and melt the right half of the supercell to obtain solid-liquid coexistence configurations. Then multiple parallel $NPT$ MD simulations (here a total of $N=N_{\mathrm{solid}}+N_{\mathrm{liquid}}$) are performed, in order to measure the probability distribution. I then infer the melting temperature based on the distribution I observed.
Right: Final states of sixteen MD duplicates, which all started from 50-50 solid-liquid coexistence. The outcome is five liquids and eleven solids, as evident from both the structures and the enthalpies (a liquid has a disordered structure and a relatively higher enthalpy). Thus $T=3100$ K, $n=16$ and $k=5$. Performing similar simulations at various temperatures allows me to collect a set of vectors (\textbf{T}, \textbf{n}, and \textbf{k}), based on which I then fit the melting temperature.
}
\end{figure*}

\section{DFT: small-size coexistence}
While being highly accurate and hysteresis-free, the traditional coexistence method is prohibitively expensive in the context of density functional formalism \cite{Kohn1964,Kohn1965}, due to the large simulation size required.
Reducing the system size would allow much faster speed, as DFT typically scales around $O(N^3)$. 
However the solid-liquid interfaces will be too close to each other and thermal fluctuations in the interface position will have a magnitude similar to the simulation cell size.
Therefore, for a small supercell, the small size renders the coexistence unstable, and the latter soon starts to form a single phase, either entirely solid or entirely liquid, which in practice fails the coexistence method in small-size applications.

I resolve this problem by introducing the small-cell coexistence approach.
%The details of the method is available in our previous paper \cite{Hon13}, which we briefly describe below.
Instead of one single long MD trajectory of a large supercell, the method runs duplicated solid-liquid coexisting simulations on small-size systems, which allow me to drastically reduce computational cost.
Melting temperatures are rigorously inferred based on statistical analysis of the fluctuations in the systems. As one type of coexistence methods, this approach also does not suffer from superheating errors (Section \ref{sec_superheating}). I have shown that finite size errors (Section \ref{sec_finite_size}) can be mostly eliminated if I choose a size sufficiently large but still feasible for DFT. 
Overall the method is highly accurate, achieving an accuracy within 100K in most materials I study based on DFT PBE calculations \cite{PBE} (Section \ref{sec_dft_error}). For certain systems where PBE errors are relatively large, I suggest making a correction based on the HSE functional \cite{HSE} (Section \ref{sec_hse}), which along with PBE often gives the lower and upper boundaries of the melting temperatures.

\subsection{Methodology}

The overall computation procedure is summarized in Fig. \ref{diagram}.
First, starting from a periodic supercell of single crystal solid, half of the supercell is heated and melted, while the atomic positions of the other half are fixed.
In the heating process, I gradually increase the temperature, until it is sufficiently high that the solid half melts, typically occurring at well above the material's melting temperature. 
This process generates an unbiased 50-50 distribution of solid and liquid composition.
After half of the supercell is melted, further MD steps are carried out while the other half is still frozen, in order to prepare multiple different configurations (or snapshots) of half-half solid-liquid coexistence.
These snapshots serve as initial starting points of $NPT$ MD duplicates (these samples differ in liquid configurations), which then undergoes MD simulations with atomic positions fully relaxed. These MD duplicates' trajectories will statistically reflect the relative phase stability of the solid and the liquid, i.e., the relative Gibbs free energy at the simulation temperature. Below melting, the two-phase coexistence is statistically more likely to solidify, and vice versa.
Thanks to the small system size, the two-phase coexistence quickly evolves to a single-phase equilibria, either entirely solid or entirely liquid, with a probability determined by the system's temperature relative to its melting temperatures.
The solid-liquid probability distributions $p_{\text{solid}}$ and $p_{\text{liquid}}$ follow the relations
\begin{eqnarray}
\frac{p_{\text{liquid}}}{p_{\text{solid}}} &=& \exp \left( - \frac{G^{l-s}(T)l_x}{2k_BT} \right), \\
H^{s/l}(T) &=& H^{s/l}(T_m)+C_{p}^{s/l}(T-T_m),\\
S^{s/l}(T) &=& S^{s/l}(T_m)+C_{p}^{s/l}\ln{\frac{T}{T_m}},\\
G^{l-s}(T) &=& G^l(T)-G^s(T) \nonumber\\
&=&\frac{(T_m-T)}{T_m}H^{l-s}(T_m) - C_p^{l-s} \frac{(T-T_m)^2}{T_m}.
\end{eqnarray}
Detailed derivation of Eqns. (1)-(4) and the validation can be found in Ref. \onlinecite{Hong2013} and are omitted here. Through fitting, I obtain melting properties, e.g., melting temperatures $T_m$, solid and liquid enthalpies $H^{s/l}(T_m)$ at $T_m$ and heat capacities $C_p^{s/l}$. Here $G$ is Gibbs free energy, $S$ is entropy, and $l_x$ is a finite-size parameter.

\begin{figure*}
\includegraphics[width=0.49\textwidth]{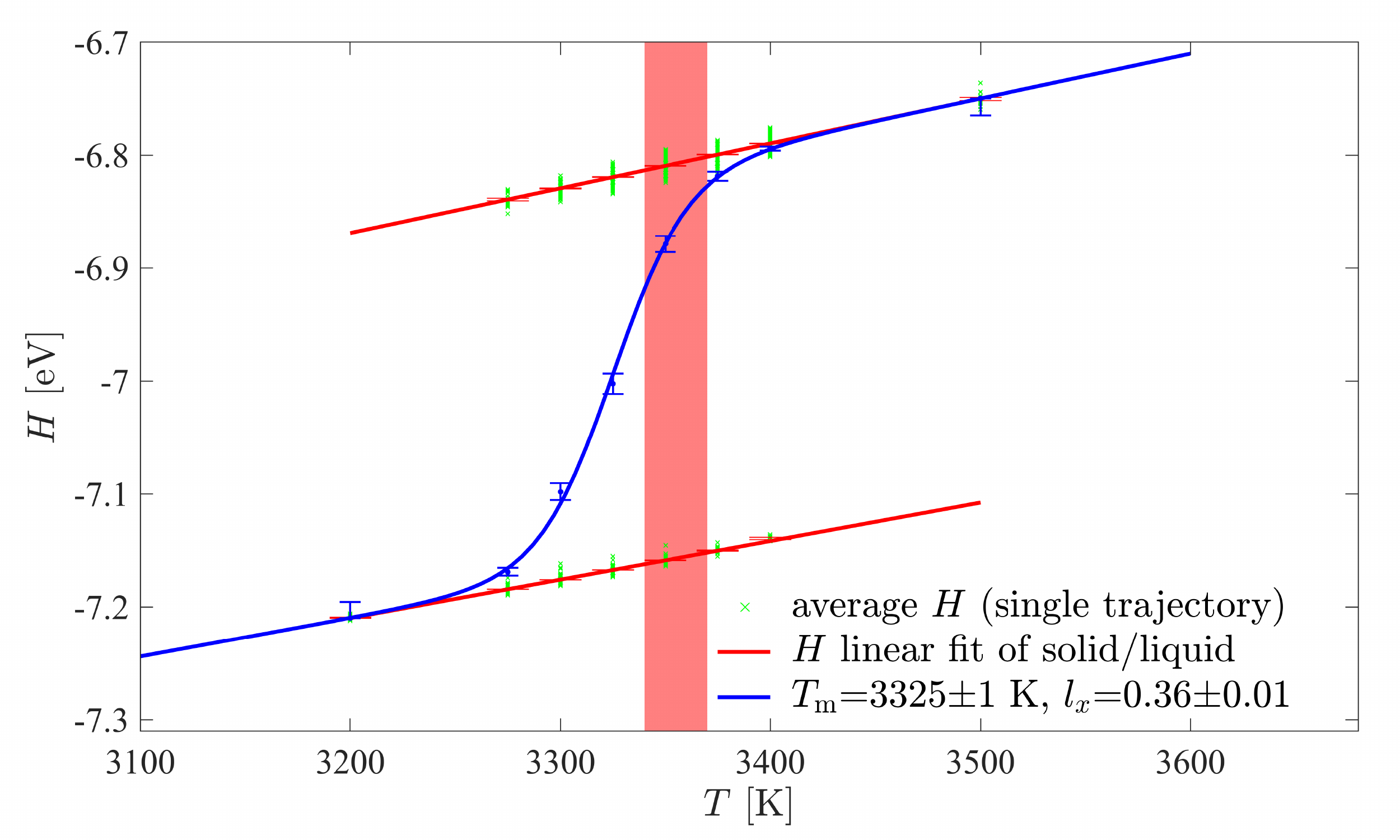}
\includegraphics[width=0.49\textwidth]{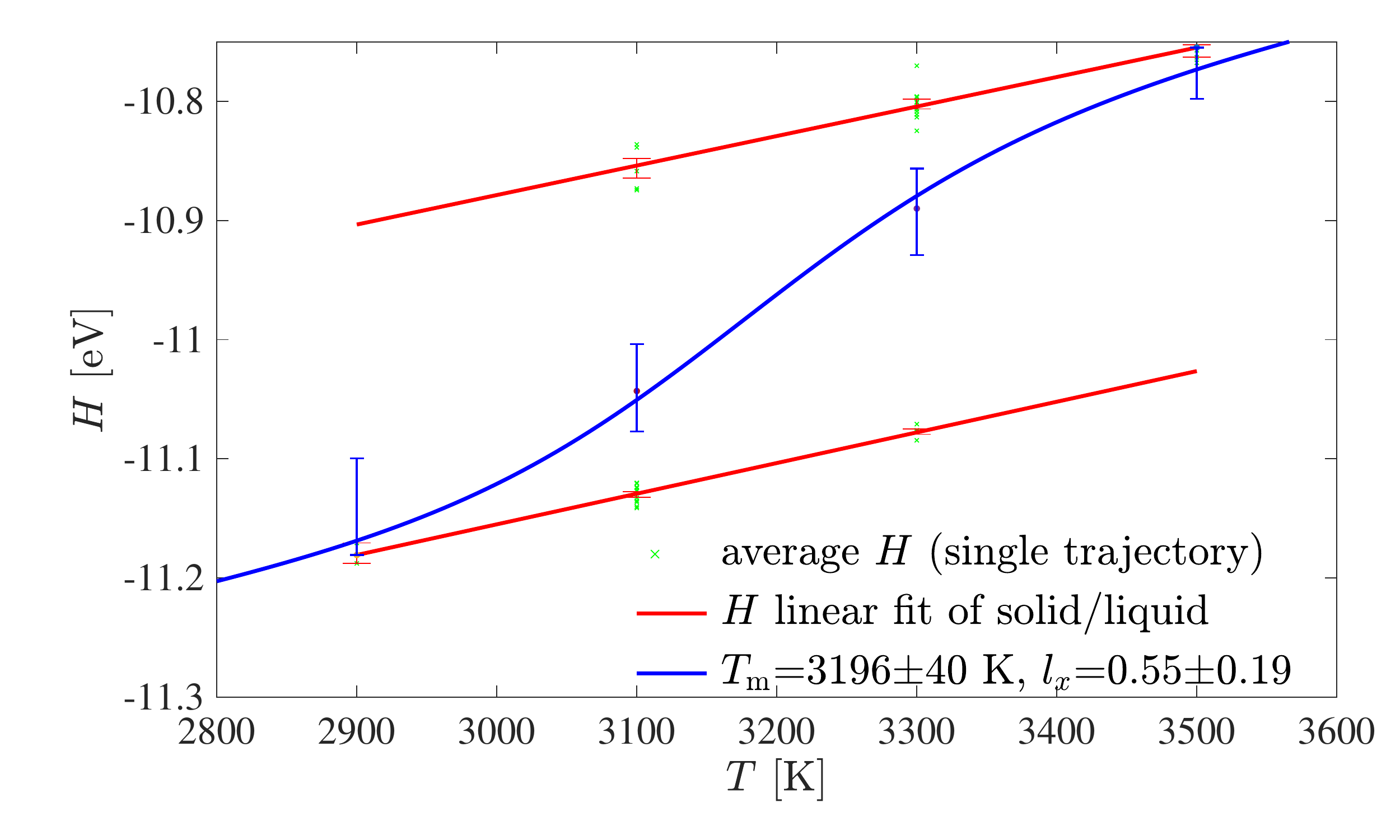}
\includegraphics[width=0.49\textwidth]{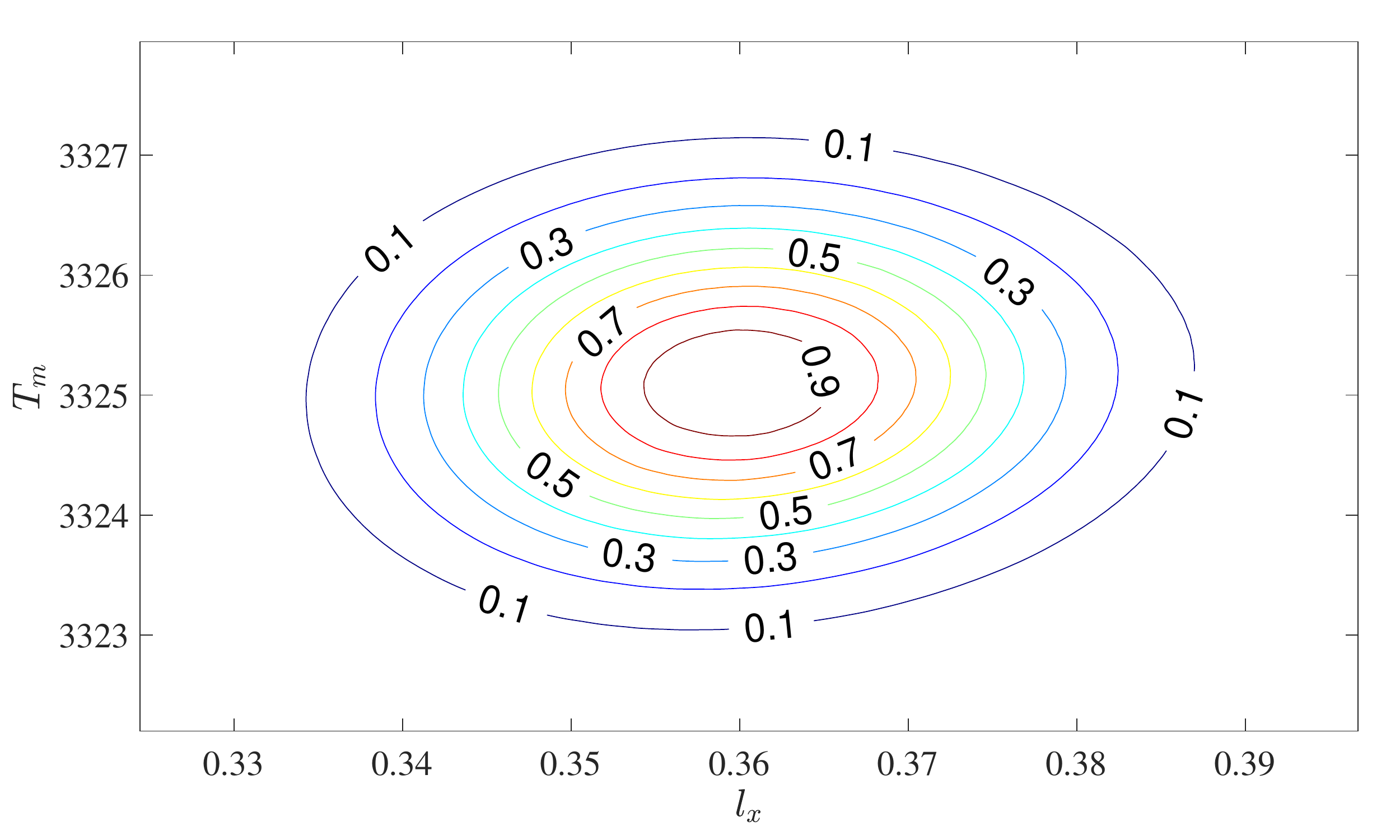}
\includegraphics[width=0.49\textwidth]{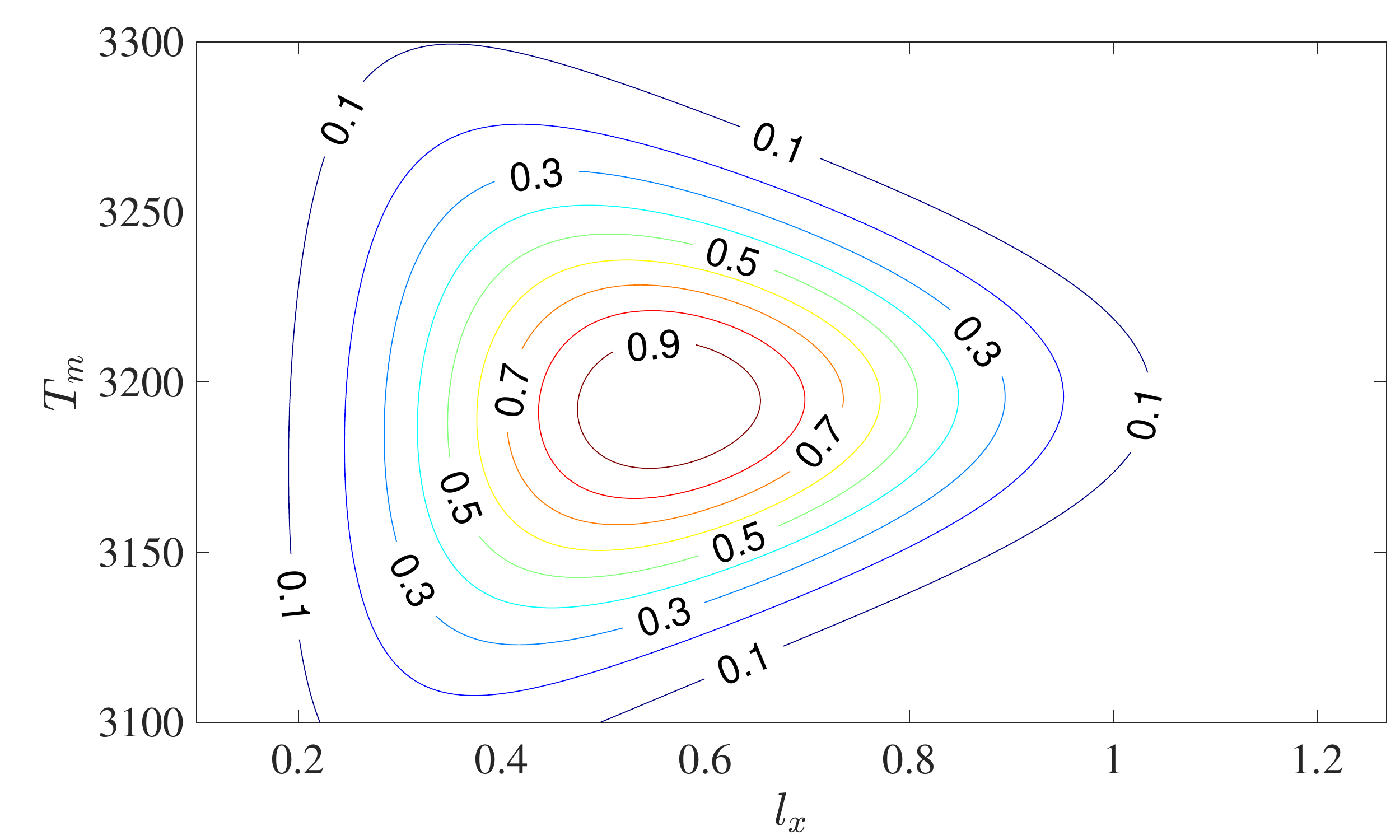}
\caption{\label{fit} Melting temperature fitting (upper) and the computed likelihood ${\cal L}\left( T_m, l_x \right)$ (lower).
(Left) Simulation and fitting were performed on a Ta empirical potential with a $6\times6\times12$ b.c.c. supercell. I was able to sample hundreds of MD trajectories with different temperatures and initial structures, achieving a high precision and a smooth curve. The calculated melting temperature is benchmarked with the standard large-size coexistence method shown as the pink bar, using the same empirical potential. 
(Right) For DFT, I carry out only a limited number (typically a few dozens) of MD simulations on a small $3\times3\times6$ b.c.c. supercell, due to the high cost. The small number of samples leads to a relatively large uncertainty of the melting temperature. However, a few dozen are sufficient to achieve a precision of 100 kelvin. 
}
\end{figure*}

In Fig. \ref{fit}, I illustrate the implementation of Eqns. (1)-(4) in melting temperature fitting. 
I denote that \textit{SLUSCHI} runs solid-liquid coexistence at multiple $s$ temperatures $\textbf{T}=\{ T_1,T_2,\cdots,T_s \}$ with $\textbf{n}=\{ n_1,n_2,\cdots,n_s \}$ samples, which yield $\textbf{k}=\{ k_1,k_2,\cdots,k_s \}$ liquids and $\{ n_1-k_1,n_2-k_2,\cdots,n_s-k_s \}$ solids, as well as their corresponding enthalpies during MD simulations. 
I plot the final enthalpies of the $\textbf{n}=\{ n_1,n_2,\cdots,n_s \}$ samples vs. the temperatures, which clearly form a binary outcome, either entirely solid or entirely liquid.
Because $C_p^{s/l}$ is the slope of enthalpy-temperature curve (assuming $C_p^{s/l}$ is independent of temperature), I obtain $C_p^{s/l}$ through two independent linear fittings on the enthalpies of the solid and the liquid, as in Fig. \ref{fit}. This leaves $T_m$ and  $l_x$ the only two independent parameters yet to determine in Eqns. (1)-(4) ($H^{l-s}(T_m)$ is a function of $T_m$).

In practice, the purpose of melting temperatures fitting is to infer the values and the statistical errors of $T_m$ and  $l_x$ from the sampled data (\textbf{T}, \textbf{n}, and \textbf{k}).
This is achieved following the principles of the maximum likelihood estimation method \cite{MLEpaper,Eli93}.
From a statistical point of view, the sampling of solid-liquid coexistence's final outcome, at one individual temperature $T$, follows a binomial distribution with parameters $n$ and $p$, i.e., the discrete probability distribution of the number of successes in a sequence of $n$ independent yes/no experiments (Bernoulli trials), each of which yields success with probability $p$. 
The probability function $f$ of getting $k$ liquids (and $n-k$ solids) in $n$ samples is 
\begin{equation}
%f\left(k,n \rvert p\right) = C_n^k p^k \left(1-p\right)^{n-k} , \text{ where } p = \frac{p_{\text{liquid}}}{p_{\text{solid}} + p_{\text{liquid}}}.
f\left(k,n \rvert p\right) = C_n^k p^k \left(1-p\right)^{n-k} , \text{ where } p = \frac{p_{\text{l}}}{p_{\text{s}} + p_{\text{l}}}.
\end{equation}
Here $p_{\text{s}}$ and $p_{\text{l}}$ are functions of the sampling temperature $T$, the melting temperatures $T_m$ and the finite-size parameter $l_x$, according to Eqns. (1)-(4). Hence
\begin{eqnarray}
f\left( k,n \rvert T, T_m, l_x \right) &=& f \left( k,n \rvert p\left(T,T_m, l_x\right) \right)  \nonumber\\
&=& C_n^k \left[ p\left(T,T_m, l_x\right) \right]^k \left[ 1-p\left(T,T_m, l_x\right)\right]^{n-k}. 
\end{eqnarray}
As I sample multiple temperatures \textbf{T}, the joint probability function for a set of vectors (\textbf{T}, \textbf{n}, and \textbf{k}) is a multiplication of probabilities at all temperatures.
\begin{eqnarray}
f \left( \textbf{k},\textbf{n} \rvert \textbf{T},T_m, l_x \right) &=& \prod_{i=1}^{s}{f_i \left( k_i,n_i \rvert T_i,T_m, l_x \right)} \nonumber \\
&=& \prod_{i=1}^{s}{f_i\left(k_i,n_i \rvert p_i \left(T_i, T_m, l_x \right) \right)}.
\end{eqnarray}

%\subsubsection{Likelihood function}

Given a set of parameters $\{T_m, l_x\}$, the probability function $ f \left( \textbf{k},\textbf{n} \rvert \textbf{T},T_m, l_x \right) $ tells how probable the data (\textbf{T}, \textbf{n}, and \textbf{k}) are. However, what I face in practice is the inverse problem: I already observe the data (\textbf{T}, \textbf{n}, and \textbf{k}), and I need to infer $\{T_m, l_x\}$, i.e., to find the parameters $\{T_m, l_x\}$ that are likely to have produced the data.

I define the likelihood function
\begin{equation}
{\cal L}\left( T_m, l_x \rvert \textbf{k},\textbf{n},\textbf{T} \right) =f \left( \textbf{k},\textbf{n} \rvert \textbf{T},T_m, l_x \right).
\end{equation}
I note the difference between the probability function and the likelihood function.
The probability function is a function defined on the data scale, given a particular set of parameter values. On the contrary, the likelihood function is a function defined on the parameter scale, given a particular set of observed data.
In practice, I create a 2-D grid of $T_m$ and $l_x$ and evaluate the likelihood function ${\cal L}\left( T_m, l_x \rvert \textbf{k},\textbf{n},\textbf{T} \right)$ at each grid point $\{T_m, l_x\}$, while the observed data (\textbf{T}, \textbf{n}, and \textbf{k}) are kept constant, as illustrated in Fig. \ref{fit}. The likelihood allows us to obtain the value and statistical error of $T_m$.

\subsection{\textit{SLUSCHI}\label{sec_sluschi}}
I have automated the computational process into the \textit{SLUSCHI} (Solid and Liquid in Ultra Small Coexistence with Hovering Interfaces) package \cite{Hong2016}, which I have freely distributed and is publicly available at my group's website. 
\textit{SLUSCHI} is fully automated, with interface to VASP. \cite{VASP} A user needs to input the crystal structure and a list of calculation parameters, and \textit{SLUSCHI} will handle the rest of the calculations, including all the simulation steps in Fig. \ref{diagram}, deciding which temperature range to sample and run simulations, and submission and analysis of all VASP jobs.
Typically it takes between several days to weeks to complete melting temperature calculation on one material.
Thanks to this automation capability, I have studied hundreds of materials by utilizing the small-size coexistence method and \textit{SLUSCHI}.

\subsection{Superheating error\label{sec_superheating}}
Superheating is the phenomenon when melting is somehow suppressed by an energy penalty, e.g., solid-liquid transition barrier, high interfacial energy, or short DFT MD time scale, resulting the solid phase to persist in its metastable solid form.
Heterogeneous melting involves solid to be partially melted into liquid, forming solid-liquid coexistence and interface, which is a large energy penalty if the cell size is relatively small.

\begin{figure}[t]
\includegraphics[width=0.49\textwidth]{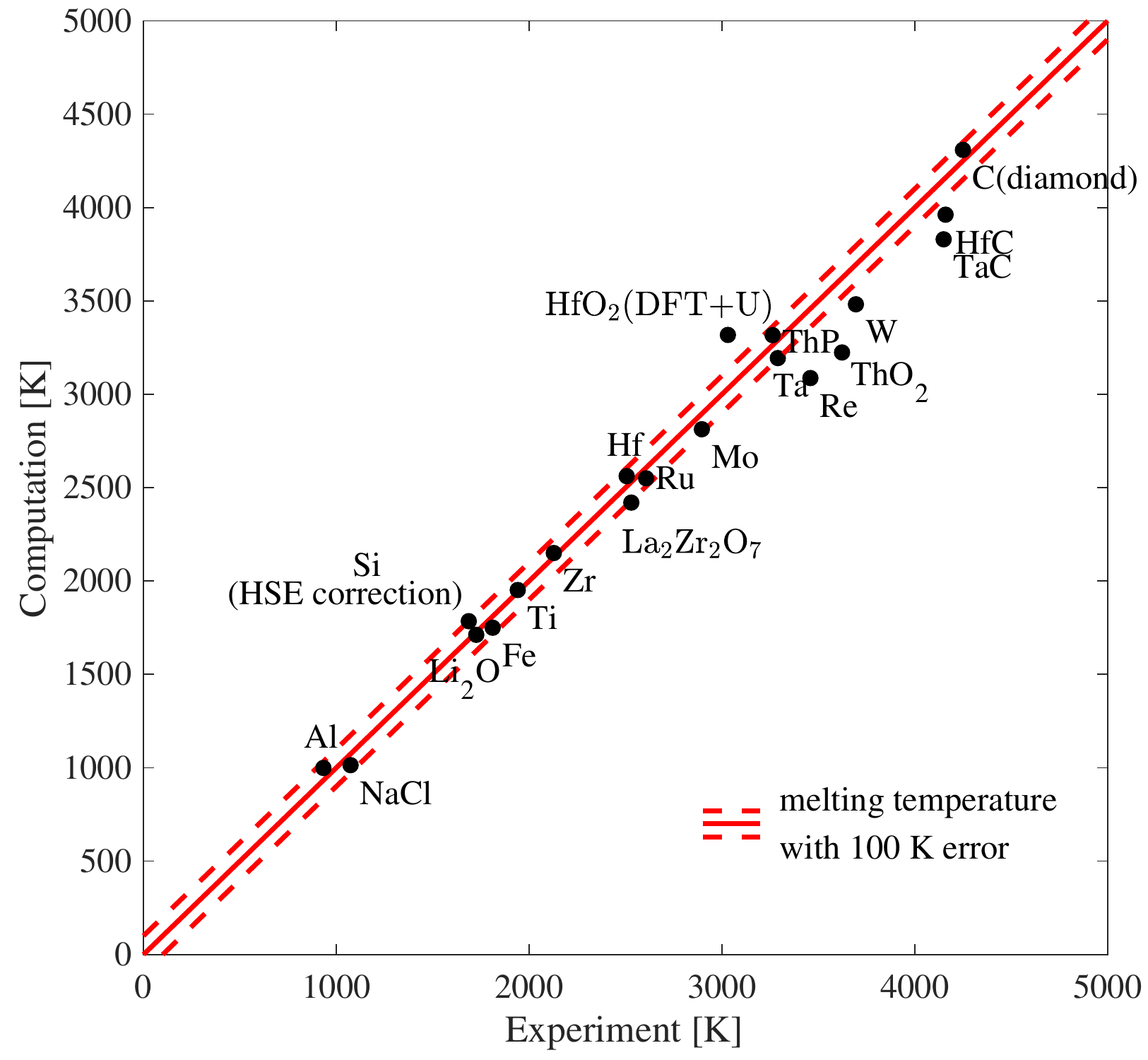}
\caption{\label{dft_summary}Computational melting temperature based on DFT versus experiments. 
Error bars of 100 K were plotted as red dash lines for eye guide.
Plotted are materials at ambient pressure that meet the requirements that (1) I have studied the materials using the method, and that (2) there exist experimental values to compare with.
The GGA-PBE functional is employed unless elaborated in the label.
PBE tends to underestimate melting temperature, most likely due to its under-binding nature.
PBE-HSE often provides lower and upper boundaries of the melting temperature.
}
\end{figure}

While methods that start from a single phase typically suffer from superheating, this type of error is mostly eliminated in solid-liquid coexistence methods, because solid-liquid interfaces are already present starting from the very beginning of the simulation, and there is no need of creating an interfaces. The small-size coexistence method and the \textit{SLUSCHI} package do not suffer from this type of error. I note that in practice \textit{SLUSCHI} PBE calculations tend to underestimate melting temperature, as suggested in Fig. \ref{dft_summary}, with no observation of superheating.

\subsection{Finite size error\label{sec_finite_size}}

Finite size error is another type of error in melting point calculations. 
MD simulations are carried out under the constraint of small periodic cells in VASP.
Under the constraint, each atom and its periodic images are required to move in the exactly same manner during MD simulations, thus limiting the degree of freedom to $3N$, where $N$ is the number of atoms in the cell. 
When the cell size is small, this limit could result in a large bias from real solid or liquid.

\begin{figure}[t]
\includegraphics[width=0.49\textwidth]{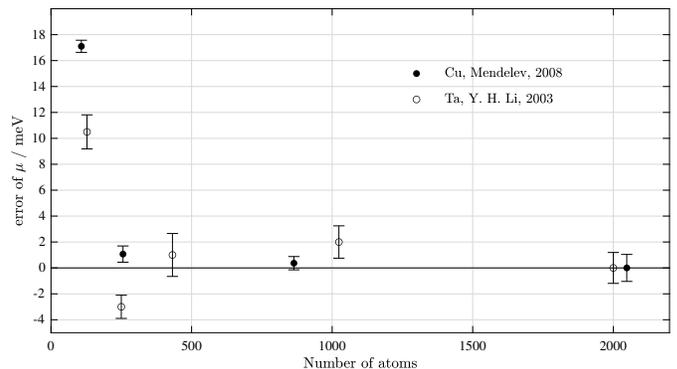}
\caption{\label{size_effect}Finite-size error of Gibbs free energy of two liquids, copper and tantalum. This study \cite{Hong2012} was carried out using the particle insertion method \cite{Widom1982} on two empirical potentials in \textit{LAMMPS}.
Even if cell size is reduced to $\sim100$ atoms, the finite size error is only 10-20 meV, which corresponds to a $\sim$100 K error in melting temperature.}
\end{figure}

At its first glance, this constraint appears worrisome, especially for the liquid phase which is naturally long-range disordered.
However, our investigation confirms that a reasonably large supercell (e.g., a size of 100 atoms) will not only be feasible in terms of DFT MD speed, but also eliminate most of the finite size error.
As shown in Fig. \ref{size_effect}, when the size is sufficiently large at $\sim100$ atoms, the finite size error is typically reduced to 10 meV/atom in terms of Gibbs free energy, which converts to approximately 100K in melting temperature calculations.

Error cancellation is another factor that works in favor of the method.
Note that, by simulating solid-liquid coexistence, the method compares the relative stability of solid and liquid, so error cancellation will further improve accuracy if the errors are towards the same direction.
Imposing periodic image constraint reduces the accessible phase space and penalizes stability for both the solid and the liquid.
Partial error cancellation is expected, though the errors are not of the same magnitude (liquid error is larger).
In addition to periodic image, lattice constant mismatch in solid-liquid coexistence will put strains on the cell and thus penalties on both phases.
The solid phase generally suffers more from this type of penalty, while the liquid structure is more flexible and less susceptible to lattice constant changes.
Combining these two mechanisms, I expect a considerable portion of error cancellation in the small-size solid-liquid coexistence method, compared to the absolute error of the entirely liquid phase, thus further reducing the finite size error.

\subsection{DFT functional error \label{sec_dft_error}}
As illustrated in Fig. \ref{dft_summary}, the method tends to underestimate melting temperature, as the PBE melting temperatures are lower than the experimental ones for most of the materials.
I believe that DFT functional error is responsible for the underestimation.
Unless additionally elaborated, melting temperature calculations in Fig. \ref{dft_summary} were carried out based on the GGA-PBE functional.
One well known issue of GGA is that it tends to underestimate bond strength \cite{Perdew2008}. Since there are more bonds in the solid phase than in the liquid, the binding energies of the solid phase are underestimated by a larger amount than those of the liquid phase, thus making the solid phase relatively unstable.
Therefore, an unstable solid phase results in an early melting, i.e., a lower melting temperature compared to experiment.

In practice, we make an HSE correction, as discussed in the next section, which always increases the melting temperature and often leads to an overestimate. Thus, PBE simulation and HSE correction can respectively serve as the lower and upper boundaries of the melting temperature.

\subsection{HSE correction \label{sec_hse}}

We evaluate the HSE functional's impact on melting temperature as
\begin{equation}
\frac{T_m^{\text{HSE}}}{T_m^{\text{PBE}}}=\frac{\Delta H^{\text{HSE}}}{\Delta H^{\text{PBE}}}\text{,} 
\end{equation}
where $\Delta H$ is heat of fusion and $T_m$ is melting temperature.
Since it is prohibitively expensive to compute $\Delta H^{HSE}$ directly, which would require MD simulation with the HSE functional,
we calculate the energy correction as a first-order perturbation, 
\begin{equation}
H^{\text{HSE}}-H^{\text{PBE}} = \left\langle H^{\text{HSE}} - H^{\text{PBE}} \right\rangle _{{\text{PBE}}} \text{.} 
\end{equation}
The bracket $\left\langle \cdots \right\rangle _{{{\text{PBE}}}}$ means that 
we randomly choose snapshots from MD trajectories of PBE, to calculate the energy differences between the two functionals.

This correction Eq. (9) assumes that enthalpy is the dominant factor in this correction, while the impact from entropy is negligible, i.e., melting temperature is proportional to heat of fusion, $\Delta H = \Delta S \cdot T_m$, while the entropy change of melting $\Delta S$ is a constant for different functionals such as PBE and HSE. 
This turns out to be a good approximation from the perspective of potential energy surface (PES).
A constant $\Delta S$ corresponds to the situation where the relative landscape of the PES remains the same, while a change in $\Delta H$ points to a shift of the PES from PBE to HSE. 
As the shifts are different for solid and liquid, the heat of fusion $\Delta H$ HSE correction is a non-zero term.
In practice, this correction often well captures the impact of different DFT functionals on melting temperature calculations, thus offering us a range of values  to serve as the lower and upper boundaries of the melting temperature.

\subsection{Summary \label{sec_sum}}
The small-size coexistence method is one of the best methods available for melting temperature prediction.
As an approach based on direct DFT MD simulations, the method is robust and flexible, suitable for a wide range of materials.
The small system size allows rapid speed, as DFT typically scales near $O(N^3)$ and this strategy drastically reduces computational cost.
In addition, the MD duplicates are perfectly paralleled, as they are separate and independent MD runs.
The method is carefully designed to reduce various types of errors. As a coexistence method, it is free of superheating error. The system size is properly selected to balance the finite size error and computational cost. While DFT functional error is inevitable, the HSE correction provides an option to generate the lower and upper boundaries of the actual melting temperature.
An attractive feature of this general approach is that it can relatively quickly deliver approximate results whose accuracy can be systematically improved by simply running the code for longer time. \vspace{.5cm}

Despite the progress, there remain challenges and hurdles to achieve rapid melting temperature prediction. At the pace of several days per material, the speed of DFT melting temperature calculation is still very limited, which puts a serious constraint on its predictive power. In the next section, I introduce an alternative approach via deep learning.\vspace{.5cm}

\section{Deep learning: graph neural networks}

In this work, I built a machine learning (ML) model to predict melting temperature from chemical formula \cite{HongML2021,HongML2022}. The design, that no other input is required in addition to chemical formula, allows me to build a model extremely easy to use for everyone in the community.
I first built a melting temperature database that contains 9375 materials. Based on the database, I built a ML model, which is capable of predicting melting temperature in milliseconds per material.
The model features graph neural network and residual neural network architecture. The root-mean-square errors of melting temperature are 75 and 138 Kelvin for training and testing, respectively. I have deployed model online and made it publicly available. A user may type in a chemical formula and receive the computational results in seconds.% In order to compute melting temperature, a user may visit my group's webpage and type in the chemical formula of the materials. Within several seconds, the ML melting temperature will be computed and displayed.

%I demonstrate potential applications of the model in materials design and discovery, as the model, without prior knowledge, quickly locates novel ternary materials of high melting points.

\subsection{Database and model}
In order to utilize ML to predict melting temperature, I first build a melting temperature database via web crawling. 
Experimental melting temperatures are collected, mostly from Ref. \onlinecite{Glushko1984}, and included in my database, and my DFT melting temperature calculations from \textit{SLUSCHI} are added as well.
My current database contains 9375 materials, out of which 982 compounds are high-melting-temperature materials with melting points above 2000 Kelvin. 
The database consists of chemical compositions, i.e., elements and concentrations of the materials, or equivalently chemical formula, and their corresponding melting temperatures.

\begin{figure}
\centering
\includegraphics[width=0.49\textwidth]{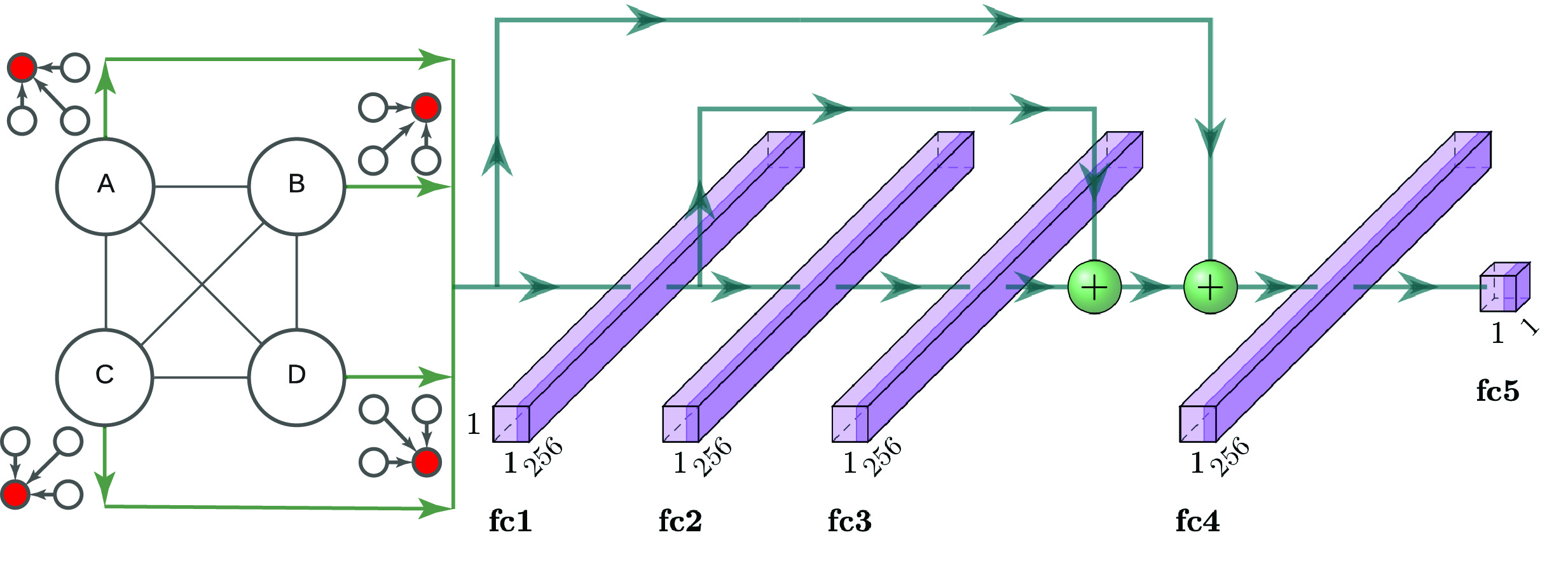}
\caption{\label{arch}Architecture of the GNN model for ML melting temperature prediction. Up to four elements and compositions (denoted as circles A, B, C and D) are connected in this graph. 
Each element and composition are first converted to features and then encoded and fed to the ResNet input layer.
The circles are connected in the GNN to exchange information in order to count for higher order contributions. In other words, each circle (element and composition) pulls information from and communicates with other circles via the GNN. 
The outputs of the GNN encoders are also fed to the ResNet input layer.
The ResNet consists of four fully connect layers (fc1-4) with skipping connections and leads to the regression analysis for melting temperature prediction.
}
\end{figure}

Based on the database, I then build a deep learning model to predict melting temperature. I employ the Graph Neural Networks (GNN) \cite{Scarselli2009} architecture, along with residual neural network (ResNet) \cite{He2016}, as illustrated in the architecture in Fig. \ref{arch}. The GNN architecture allows me to incorporate physics into the neural network connections. This particular design of GNN architecture imposes permutation invariance of chemical formula, which drastically reduces training complexity and improves efficiency. The ResNet architecture effectively handle the issue of diminishing gradient in model training.

When a material (its elements and compositions) is fed to the neural network, each element that constitutes the material is first converted to 14 features, such as atomic radius, electronic negativity, electron affinity, valence electrons, position in the periodic table, etc. These features are encoded and passed to the next layer, a process considered as the individual contribution to the melting temperature from each element.
In addition, elements communicate to each other through the network connections in the GNN architecture, thus leading to impacts from the binary and ternary (and higher-order) combination of the elements. These encoded impacts are fed to the next layer as well.
This layer, consisting of singular, binary, and ternary interaction of the elements and compositions of the material, then goes through a 4-layer ResNet, which leads to the regression and the estimation of melting temperature.
Currently the number of elements are limited to four (i.e., up to quaternary compounds), in order to control model complexity and reduce overfitting. However I can increase this number in case of more elements.
The GNN architecture undergoes two iterations of network communication among elements, as we find more rounds do not further improve performance.

The 9375 materials are randomly assigned to training and testing sets, with 8635 materials in the training set, and 740 materials for testing. 
The neural network model is built with the Tensorflow system \cite{tensorflow}. The training process takes approximately 2000 epochs of optimization.
The root-mean-square errors (RMSE) of melting temperature are 75 and 135 K for the training and testing sets, respectively.

\subsection{Deployment}
We deploy the model at my group's website \cite{ML_model_webpage}, so other researchers, especially those without the knowledge of machine learning, may use the Application Programming Interface (API) to estimate the melting temperatures for the materials of their interests. 
The model is currently hosted at Microsoft Azure and the ASU Research Computing facilities. 
To use the model, a user may visit our group's website and input the chemical formula. After submission, the model will respond in seconds with the predicted ML melting temperature, as well as the actual experimental melting temperatures of the ``nearest neighbors", i.e., the most similar materials, in the database. Thus this model serves as not only a predictive ML model, but a handbook of melting temperature as well.
A user may also run bulk calculations via command line with much shorter latency, by sending an HTTP POST request to the API server and providing JSON data (that contains elements and compositions of multiple materials) in the body of the POST message. Detailed instruction is available at the website.

\subsection{Summary}

\begin{figure}
\includegraphics[width=0.49\textwidth]{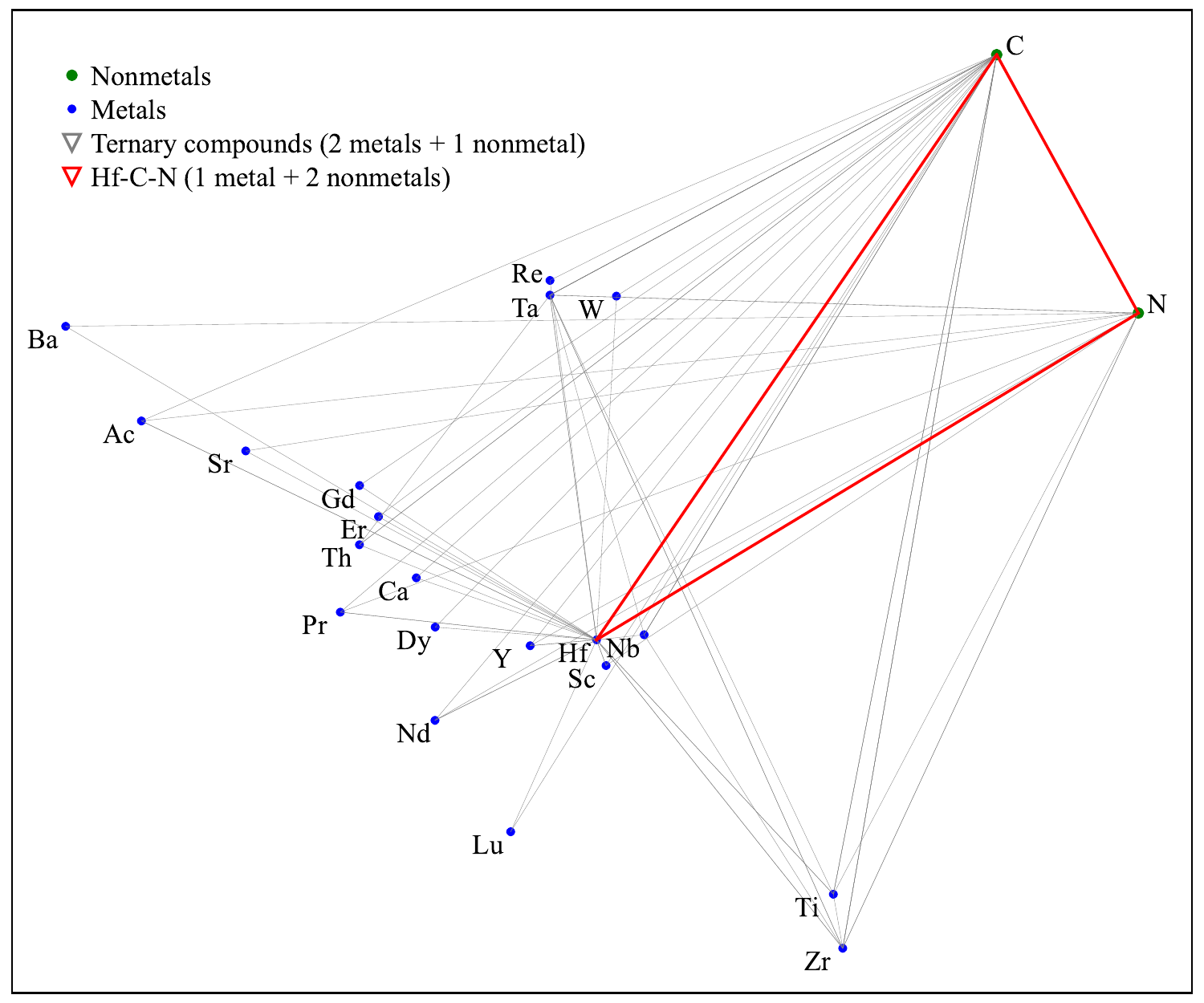}
\caption{\label{ternary} Prediction of high melting temperatures in ternary systems, based on the GNN model and Monte Carlo simulated annealing. 
Shown in the figure are top ternary compounds that have the highest melting temperatures.
Each triangle represents one ternary compound, with the three corners being the elements.
The top candidates are dominated by carbides and nitrides. The only exception is a carbonitride, the Hf-C-N system, which is exactly the material of the highest melting temperature we predicted in 2015 based on DFT MD simulations \cite{HfCN2015}. Note that the DFT melting temperatures of Hf-C-N are excluded in the ML model, in order to test its predictive capability.}
\end{figure}

At first glance, this model appears remarkably accurate, in addition to its extremely rapid speed, compared to other complex and expensive methods, such as DFT calculations.
However I would like to point out that, as a machine learning model, while it may perform well in certain composition space, which is typically well sampled, it may fail in under-explored space, where the model relies on extrapolation.
In the latter scenario, DFT will provide robustness and reliability, which not only complements the GNN model but also expands its dataset into new composition space for learning.
The small-size coexistence method and \textit{SLUSCHI} perfectly fit in this role, as the automated tool generates accurate DFT melting temperatures.
A larger dataset that covers a broader chemical composition space will further improve the accuracy of the GNN model.
An integration of the GNN model and the DFT tool will allow me to build an iterative and self-consistent framework for melting temperature prediction and materials design and discovery of high melting temperature materials.
We can also utilize the GNN model, along with approaches such as simulated annealing as demonstrated, to solve the optimization problem of melting temperature maximization, as illustrated in Fig. \ref{ternary}.

To summarize, I have built a melting temperature database and a machine learning model for melting temperature prediction. 
The database so far contains approximately 10,000 materials, and the GNN model achieves high accuracy with 75 and 138 K root-mean-square error of melting temperature for training and testing, respectively.
The model is deployed online and publicly available.
Future integration of the database, the ML model, and the DFT tool will create a framework for melting temperature prediction and materials design and discovery based on melting temperature related properties. \vspace{.5cm}

In this work, I reviewed two methods I have built over the past few years, with an ultimate goal of melting temperature prediction.
The methods are complementary to each other: the deep learning model is extremely fast and easy to use, while \textit{SLUSCHI} is highly accurate and robust, with a solid track record of application to a wide range of materials.
An integration of DFT and deep learning in the future will bring within reach our goal of accurate and rapid prediction of material's melting temperature.

\section*{Acknowledgements}
This research was supported by Arizona State University through the use of the facilities at its Research Computing. This work uses the Extreme Science and Engineering Discovery Environment (XSEDE), which is supported by National Science Foundation Grant No. ACI-1548562.
Hong would like to thank Axel van de Walle for his advices and support. 

\bibliography{citepapers}

\end{document}